\documentclass[doublecol]{epl2} 
\usepackage{epsf}
\usepackage{graphicx}
\usepackage[justification=justified]{caption}
\usepackage{amsmath}
\usepackage{tabularx, booktabs, makecell}

\title{Spectroscopic evidence for  temperature--dependent convergence of light and heavy hole valence bands of PbQ (Q=Te, Se, S)}
\shorttitle{Temperature--dependence of light and heavy hole valence bands of PbQ (Q=Te, Se, S)} 

\author{J. Zhao\inst{1},  C. D. Malliakas\inst{2,3}, K. Wijayaratne\inst{1}, V. Karlapati\inst{1}, N. Appathurai\inst{1}, D. Y. Chung\inst{3}, S. Rosenkranz\inst{3}, M. G. Kanatzidis\inst{2,3} \and U. Chatterjee\inst{1}}
\shortauthor{J. Zhao \etal}

\institute{                    
  \inst{1} Department of Physics, University of Virginia - Charlottesville, VA 22904, USA\\
  \inst{2} Department of Chemistry, Northwestern University - Evanston, IL 60208, USA\\
  \inst{3} Materials Science Division, Argonne National Laboratory - Argonne, IL 60439, USA
}

\abstract{
We have conducted temperature dependent Angle Resolved Photoemission Spectroscopy (ARPES) study of the electronic structures of PbTe, PbSe and PbS. Our ARPES data provide direct evidence for the  \emph{light} hole upper valence bands (UVBs) and hitherto undetected \emph{heavy} hole lower valence bands (LVBs) in these materials. An unusual temperature dependent relative movement between these bands leads to a monotonic decrease in the energy separation between their maxima with increasing temperature, which is referred as band convergence and has long been believed to be the driving factor behind extraordinary thermoelectric performances of these compounds at elevated temperatures.}

\begin{document}

\maketitle

\section{Introduction}
Lead chalcogenides PbQ (Q=Te, Se, S) are canonical systems for fundamental studies of thermoelectric (TE) properties \cite{Vineis2010, Snyder2012, Dresselhaus2007} due to their unique electronic structures. Recently, new concepts of ``all scale hierarchical architecture processing" \cite{Kanatzidis2014,Kanatzidis2010, Snyder2011} have lead to significant advancements in their TE performance. For instance, $p$-type nanostructured PbTe holds the current performance record for high temperature energy conversion \cite{Girard2011, Pei2011, Biswas2011b}. Despite being studied for  decades PbQ consistently surprise us with new findings. One such example is the recently discovered appearance of local Pb off-centering dipoles on warming without a structural transition in PbTe  \cite{Bozin2010, PbTe_Sales}. Moreover, these systems have recently been shown to host  various novel quantum states of matter. For instance, Pb$_{1-x}$Sn$_x$Se, Pb$_{1-x}$Sn$_{x}$Te are shown to be topological crystalline insulators \cite{Liang Fu, Hasan_PbTe, Ando_PbTe, Dziawa_PbSe, Vidya_PbSe, TCI_N_1, TCI_N_2, TCI_N_3, TCI_N_4}, while superconductivity along with normal state charge Kondo anomaly occurs at Tl doped PbTe \cite{Ian Fisher}. 



In a number of reports, $T$ dependent thermopower of PbQ has been interpreted in terms of a relative shift with increase in $T$ \cite {Heremans2013, Snyder2013, Pei2011, Biswas2011b, UVB_LVB_2, Allgaier1968} between two different valence bands---namely, the upper valence bands (UVBs) with maxima at  L points and the lower valence bands (LVBs) due to secondary valence bands presumably occurring along $\Gamma$-K  and  $\Gamma$-X lines with maxima at lower energies compared to the UVBs \cite{UVB_LVB_1, UVB_LVB_2, UVB_LVB_3}. 
Despite similarity in the findings of these studies, there is a marked disagreement among the reported values of the crossover temperatures, i.e., $T$'s at which separations between maxima of UVBs and LVBs vanish. For example, early works going back to the 1960s and also some of the later ones concluded a crossover $T$$\sim$450K in PbTe, while it has recently been shown to be much higher $\sim$750K \cite {Heremans2013, Snyder2013}. More importantly, there is no direct experimental evidence for $T$ dependent changes in valence bands. Furthermore, there are recent alternative  descriptions to two-band analysis of thermopower data \cite{Ekuma2012, DJS}. Given all these, an in-depth examination of the electronic structures of these compounds as a function of $T$ using ARPES is highly desirable.

Recently, there have been a number of important ARPES works \cite{OLD_ARPES, NAKAYAMA, Hasan_PbTe, Ando_PbTe, Dziawa_PbSe}, on PbTe and PbSe, which are predominantly focussed on topological aspects of their electronic structures. 
In this article, employing $T$ dependent  ARPES measurements on PbQ, we resolve a lingering issue in the field: How does rising $T$ impact their valence bands? Here, we show: (i) there are two distinct valence band maxima, separated in energy as well as in momentum,  and (ii) the energy separation between these maxima decreases with increasing $T$ up until our highest measured $T$'s.

\section{Experiments}
We have carried out $T$ dependent ARPES experiments on various $n$- and $p$- type PbQ single crystal samples at the PGM beamline of Synchrotron Radiation Center, Stoughton, Wisconsin using a Scienta R4000 electron analyzer. In this work, we present ARPES data from  (a) one $n$- and  two $p$- type PbTe samples (referred as PbTe1, PbTe2, (b) two $n$-type PbSe samples (referred as PbSe1, PbSe2), and (c) two $n$-type PbS (PbS1, PbS2) samples.
ARPES measurements were performed using plane polarized light with 22 eV photon energy and data were collected at 2 or 4 meV energy intervals. The energy and momentum resolutions were approximately  20 meV and 0.0055 $\AA^{-1}$  respectively. PbQ samples were prepared by melting mixtures of Pb and Q at 100$-$150K above the individual melting points of PbQ inside evacuated fused silica tubes. PbI$_2$ was used for achieving  $n$-type doping, while Na for $p$-type. Typical carrier concentrations of the $n$ and $p$-type samples ranged from 2$-$5$\times10^{19}$cm$^{-3}$ and 0.2$-$2$\times10^{19}$cm$^{-3}$ respectively. These samples were cleaved {\it in situ} to expose fresh surface (001) of the crystal for ARPES measurements. 
\begin{figure}
\centering
\includegraphics[width=0.48\textwidth]{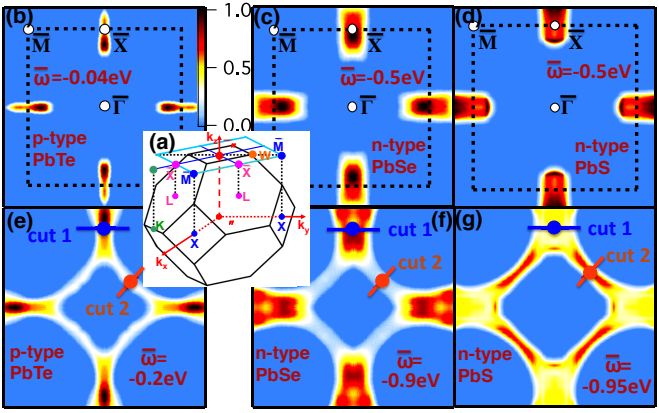}
\caption{(a) Bulk BZ (black lines) associated with a fcc crystal structure and corresponding (001) surface BZ (sky-blue lines). CEIMs of a $p$-type PbTe sample at (b) $\overline{\omega}$ = -40 meV and (e) $\overline{\omega}$ = -200 meV. (c, f) are analogous plots for $n$-type PbSe at $\overline{\omega}$ = -500 meV and -$900$ meV respectively, while (d, g) for $n$-type PbS at $\overline{\omega}$ = -500 meV and -950 meV respectively. Blue  lines in (e),(f) and (g) denote Cut1 along which ARPES data have been taken for Figs. 2(b), 2(f) and 2(j), while red lines correspond to Cut2 along which ARPES data have been taken for Figs. 2(d), 2(h) and 2(l). Blue and red dots correspond to the momentum locations of the top of the UVBs and LVBs respective. }
\label{FIG1.pdf}
\end{figure}

\section{Results}
\begin{figure*}[t]
\centering
\includegraphics[width=5.30in]{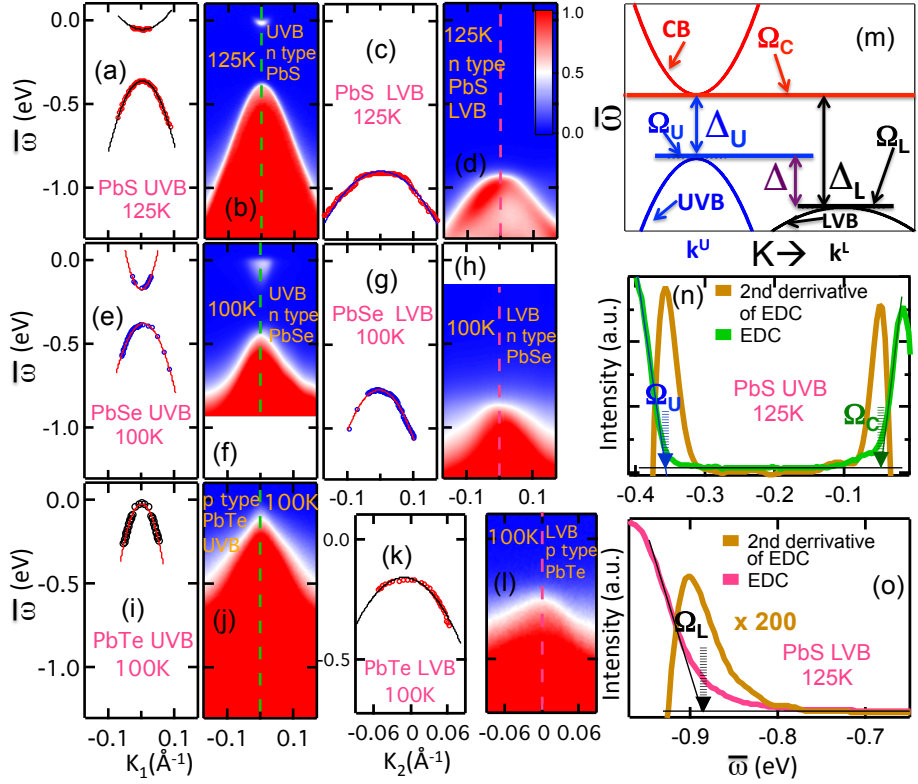}
\caption{Band dispersions (shown by open circles) and EMIMs along cut 1 for (a, b) PbS, (e, f) PbSe 
and (i, j) PbTe, and along cut 2 for (c, d) PbS, (g, h) PbSe and (k, l) PbTe. 
Parabolic fittings of the dispersions are shown by solid curves in  (a), (c), (e), (g), (i), and (k). 
(m) A schematic diagram showing relative positions of the conduction 
band (CB), LVB and UVB in PbQ. (n) EDC (green curve) and  its second derivative (golden curve) at $\bf{k}^{U}$ for PbS. The blue
line through data points in (n) guides almost linearly increasing intensity of the 
EDC with $\overline{\omega}$ for $\overline{\omega}$$<$$\Omega_{\rm U}$,  while the green line for $\overline{\omega}$$>$$\Omega_{\rm C}$. The positions of $\Omega_{\rm U}$ and $\Omega_{\rm C}$ agree well with the peak positions 
of the second derivative of the EDC. (o) EDC (magenta curve) and its second 
derivative (golden curve) at $\bf{k}^{L}$. 
Black line through data points show linear increase in intensity of the 
EDC with $\overline{\omega}$ for $\overline{\omega}$$<$$\Omega_{\rm L}$.The position of $\Omega_{\rm{L}}$ also matches nicely with the peak position of the second derivative 
of the EDC. $k_1$=$k_x$-$({\bf k^U})_x$, $k_2$={\bf k}$\cdot$${\bf \hat {e}}$ - ${\bf k^L}$$\cdot$${\bf \hat{e}}$, where $\bf{\hat e}$ is the unit vector along momentum space cut 2.
}
\label{FIG2.pdf}
\end{figure*}
In order to elucidate  the ARPES data, we first consider the bulk Brillouin Zone (BZ) of  PbQ, which has a face centered cubic (fcc) crystalline structure. This BZ is represented in Fig. 1(a).  PbQ is preferentially cleaved along (001) plane and hence, we concentrate on  its square surface BZ projected onto the (001) plane. As shown in Fig. 1(a), the $\Gamma$ and L points project on the  ($\overline{\Gamma}$) and $(\overline{\rm X})$ of the surface BZ. In Figs. 1(b)$-$1(g), we show constant-energy  intensity maps (CEIMs), i.e., ARPES data as a function of in-plane momentum components ${k_{x}}$ and ${k_{y}}$ at fixed  $\overline{\omega}$'s, where $\overline{\omega}$ is electronic energy with respect to chemical potential. Figs. 1(b) and 1(e) correspond to the CEIMs at $\overline{\omega}$ = -40 meV and -200 meV respectively for a $p$- type PbTe sample (PbTe1). Hole pockets derived from the UVBs centered at $L$ points are clearly visible around $\overline{\rm X}$  at $\overline{\omega}$ = -40 meV (Fig. 1(b)), while tubular regions connecting these isolated pockets appear at $\overline{\omega}$ = -200 meV (Fig. 1(e)). Qualitatively, similar evolution of the CEIMs with $\overline{\omega}$ can be seen for PbSe in Figs. 1(c) and 1(f), and  for PbS in Figs. 1(d) and 1(g).  In first principles calculations \cite{DJS, SVANE, FS_PBTE}, such tubular regions \cite{DJS, SVANE, FS_PBTE}  have been associated with the LVB, whose maximum occurs along the line between the $\Gamma$ and K points while other maxima occur at W and along $\Gamma$-X. Furthermore, these maxima have been predicted to lead to a quasicubic filament of valence states at higher binding energies. The $\overline{\omega}$ dependence of CEIMs in Fig. 1 seems to be in accord with these.
The method for constructing the CEIMs in Fig. 1 is as follows: starting from the raw ARPES data, we first subtract the constant signal at $\overline{\omega}$$>$0 (due to second order light) and then, we normalize each ARPES spectrum by the area enclosed by it and the energy axis between measured values of $\overline{\omega}$. The raw data covered more than one half of the surface BZ for each sample. 
For better visualization, the CEIMs in the entire BZ ware reconstructed by reflections, using interpolations to uniform grids. 

To further investigate the valence band structure of PbQ, we look into Fig. 2. Here band dispersions and ARPES energy-momentum intensity maps (EMIMs) along two specific momentum space cuts, namely cut 1 and cut 2 defined in Fig. 1, are presented. Figs. 2(b), 2(f), 2(j) correspond to EMIMs along cut 1, while Figs. 2(d), 2(h), 2(l) to those along cut 2. Figs. 2(b), 2(d) correspond to a $n$-type PbS sample, Figs. 2(f), 2(h)  to a $n$-type PbSe, and  Figs. 2(j), 2(l) to a $p$-type PbTe sample. One can recognize an electron band, i.e., the conduction band (CB), separated in energy and momentum from a hole band, i.e., the UVB, in EMIMs along cut 1 for $n$-type PbS and PbSe samples. The CB of the $p$-type PbTe sample is invisible as it is in the unoccupied side. On the contrary, a separate hole band alone is present along cut 2 in each of Figs. 2(d), 2(h), 2(l). As to the first principle calculations, the secondary valence band maxima are located away from $\Gamma$-X where the fundamental gap is found so that the next maxima would correspond to the LVB. Therefore, we  identify the hole bands along cut 2 as the so-called LVBs since their maxima are located away from $\overline{\Gamma}$-$\overline{\rm X}$. We would like to point out that  all high symmetry lines including  $\Gamma$-X, $\Gamma$-K, K-W and W-X project onto $\Gamma$-M in the (100) surface Brillouin zone. It will be an important topic for future studies to establish association of this second valence band to a specific bulk electronic state via a detailed study of the three-dimensional electronic structures. Nevertheless, Figs.1 and 2 together corroborate the two-band picture for  the valence band structure of PbQ.

\begin{table}
\caption{Values of effective mass of the UVB ($m^{*}_{UVB}$), and that  of the LVB ($m^{*}_{LVB}$) in units of electronic mass $m_e$ }\vspace{0.05in}
\centering
\renewcommand{\arraystretch}{1}
\begin{tabularx}{0.48\textwidth}{>{\centering\arraybackslash}X >{\centering\arraybackslash}X >{\centering\arraybackslash}X >{\centering\arraybackslash}X}\hline\hline 
Materials & {$\frac{m^{*}_{UVB}}{m_{e}}$}& {$\frac{m^{*}_{LVB}}{m_{e}}$}\\
\hline
PbTe & 0.045
 & 0.121
 \\   PbSe & 0.087
 & 0.141
 \\   
PbS & 0.093
 & 0.349
 \\    \hline\hline
\end{tabularx}
\label{table1}
\end{table}
Now we discuss the methodology for extracting band dispersions and band edges of UVBs and LVBs of PbQ from their EMIMs. As to the procedural details, we refer to Figs. 2(n) and 2(o), where we display Energy Distribution Curves (EDCs) along with their second derivatives with respect to (w.r.t) $\overline{\omega}$ at the $\bf{k}^{U}$, the momentum location associated with the band maximum (BM) of the UVB, and at ($\bf{k}^{L}$), the momentum location of the BM of the LVB, respectively for a PbS sample. In this context, an EDC is the distribution of electrons as a function of $\overline{\omega}$ at a fixed momentum. On closer inspection Fig. 2(n) reveals a rather sharp rise in intensity of the EDC on top of a nearly flat background signal below a certain value of $\overline{\omega}$. We define this particular $\overline{\omega}$ as $\Omega_{\rm U}$, which marks the $\overline{\omega}$ location of the BM of the UVB (Fig. 2(m)). The intensity of the EDC in Fig. 2(n) also displays similar abrupt increase above a certain value of $\overline{\omega}$. We call this $\overline{\omega}$ as $\Omega_{\rm C}$, which denotes the $\overline{\omega}$ location of the band bottom (BB) of the CB. Furthermore, $\overline{\omega}$  location of the BM of the LVB, referred as $\Omega_{\rm L}$, can be obtained by utilizing identical analysis of the EDC at $\bf{k}^{L}$  in Fig. 2(o). It is straightforward to realize that $\Omega_{\rm U}$, $\Omega_{\rm C}$ and $\Omega_{\rm L}$ also correspond to the local maxima of the second derivatives of the relevant EDCs w.r.t  $\overline{\omega}$.  This is demonstrated via Figs. 2(n), 2(o) and further elaborated in Figs. S1 and S2 of the supplementary section. As to the dispersion of an energy band, it can be tracked by detecting the local maxima of the intensity profiles as a function of $\overline{\omega}$ at a fixed momentum for a series of momentum values in the second derivative of an EMIM. Band dispersions, constructed via such second derivatives analysis, of UVBs are presented in Figs. 2(a), 2(e), 2(i), while those of LVBs in Figs. 2(c), 2(g), and 2(k). From parabolic fitting of a dispersion in the vicinity of its BM, corresponding effective mass can be approximated. Listed values  of effective masses  in Table \ref{table1} clearly show that LVBs are heavier than UVBs in PbQ. This can readily be inferred by visual comparisons of their EMIMs as well.

\begin{table}[h]
\caption{Values of $\Delta_{\rm L}$, $\Delta_{\rm U}$, and $\Delta$ obtained from $n$-type PbTe, PbSe, and PbS samples at $T$$\sim$100K}\vspace{0.05in}
\centering
\renewcommand{\arraystretch}{1.5}
\begin{tabularx}{0.48\textwidth}{>{\centering\arraybackslash}X >{\centering\arraybackslash}X >{\centering\arraybackslash}X >{\centering\arraybackslash}X}\hline\hline
Materials & {$\Delta_{\rm U}$ (meV) }& {$\Delta_{\rm L}$ (meV) }& {$\Delta$ (meV)}\\
\hline
PbTe & 190 & 320 & 130\\  
PbSe & 204 & 594 & 390\\   
PbS & 309 & 838 & 529\\    \hline\hline
\end{tabularx}
\label{table2}
\end{table}

The band gap $\Delta_{\rm U}$ of the UVB is given by:  $\Delta_{\rm U}$=$\Omega_{\rm C}$-$\Omega_{\rm U}$ (Fig. 2(m)). Similarly, the band gap $\Delta_{\rm L}$ is:  $\Delta_{\rm L}$=$\Omega_{\rm C}$-$\Omega_{\rm L}$ (Fig. 2(m)). In Table \ref{table2}, we display $\Delta_{\rm U}$ and $\Delta_{\rm L}$ for various $n$- type PbQ samples. Data from PbSe and PbS are shown in Fig. 2, while those from $n$- type PbTe sample has been displayed  in Fig. S3 of the supplementary section. The samples under current studies  do not have the exact same carrier concentration and thus, the useful quantity to be compared is $\Delta$=$\Delta_{\rm L}$-$\Delta_{\rm U}$, i.e., the difference between energy gaps of the LVB and UVB. Table \ref{table2} shows that $\Delta$ is largest for PbS and smallest for PbTe, while in between for PbSe at $T$$\sim$100K.

\begin{figure}[h]
\centering
\includegraphics[width=0.48\textwidth]{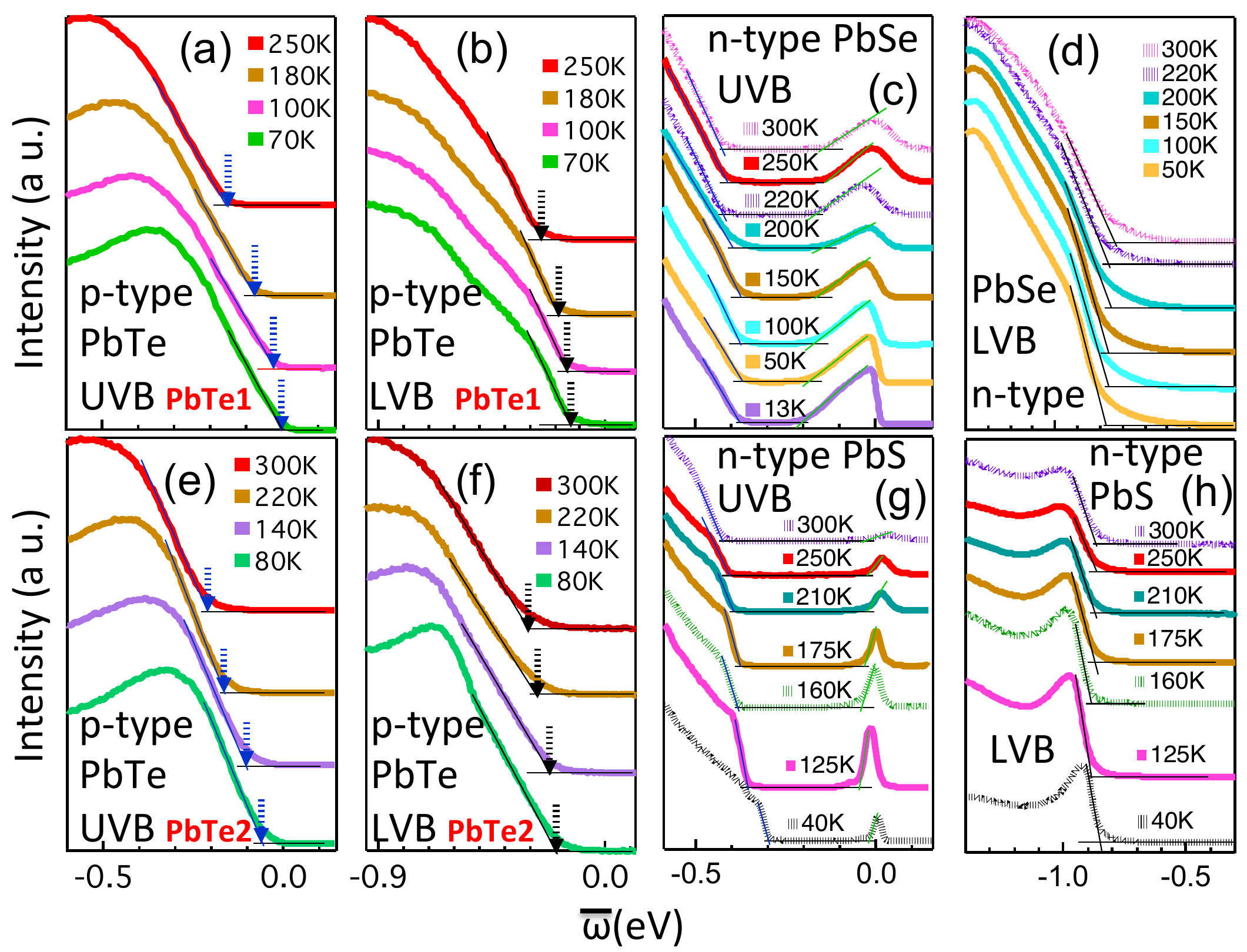}
\caption{EDCs as a function of $T$ at $\bf{k}^{U}$ for (a, e) $p$- type PbTe (PbTe1 and PbTe2), (c) $n$-type PbSe and (g) $n$-type PbS, while those at $\bf{k}^{L}$ for for (b, f) $p$-type PbTe (PbTe1 and PbTe2), (d) $n$-type PbSe and (h) $n$-type PbS. Solid and dashed lines in (c) and (d)  correspond data from PbSe1 and PbSe2 respectively, while in (g) and (h) from PbS1 and PbS2 respectively. Blue lines through data points in (a), (e), (c) and (g) exhibit rather sharp changes in slope of the EDCs at $\overline{\omega}$$\sim$$\Omega_{\rm U}$, while green lines in (c) and (g) mark those at $\overline{\omega}$$\sim$$\Omega_{\rm C}$.  Similarly, black lines through data points in (b), (d), (f) and (h) denote rather abrupt changes in slope of the EDCs at $\overline{\omega}$$\sim$$\Omega_{\rm L}$. Blue arrows in (a) and (e) point $\Omega_{\rm U}$'s, while black arrows in (b) and (f) $\Omega_{\rm L}$'s.}
\label{FIG3.pdf}
\end{figure}

\begin{figure}[h]           
\centering                                                                                                                                                                               
\includegraphics[width=0.48\textwidth]{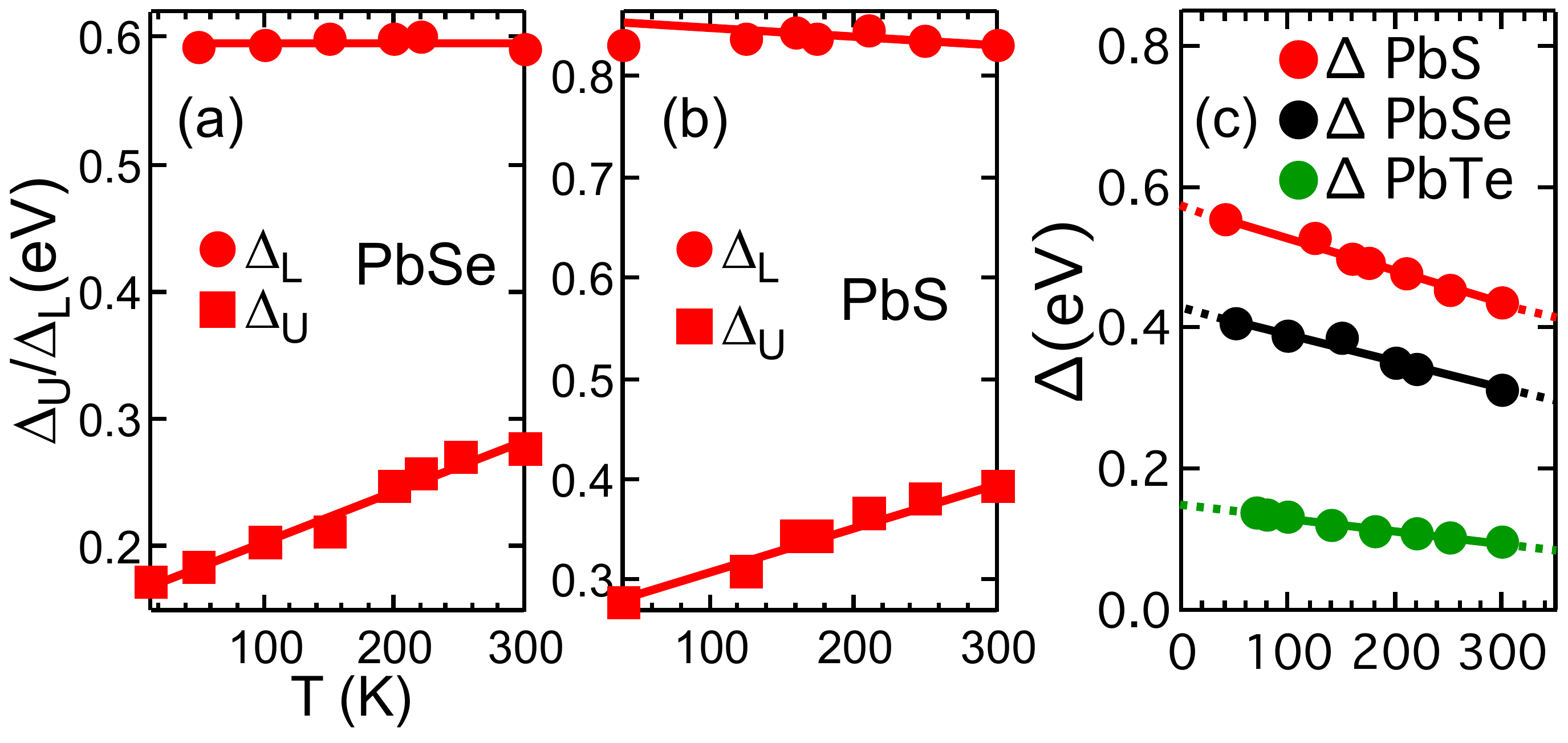}
\caption{$T$ dependence of $\Delta_{\rm U}$(=$\Omega_{\rm C}$-$\Omega_{\rm U}$) and $\Delta_{\rm L}$(=$\Omega_{\rm C}$-$\Omega_{\rm L}$) for (a) PbSe and (b) PbS. (c) $\Delta$(=$\Omega_{\rm U}$-$\Omega_{\rm L}$) as a function of $T$ for PbSe, PbS and PbTe. }
\label{FIG4.pdf}
\end{figure}
The objective of the remainder of the paper is to interrogate  the impact of increasing $T$ on valence bands of PbQ. This is summarized in Figs. 3 and 4. In Figs. 3(c) and 3(g), we plot $T$ dependent EDCs for $n$-type PbSe and PbS samples respectively at their individual $\bf{k}^{U}$s, while in  Figs. 3(d) and 3(h) at their $\bf{k}^{L}$s. We also display similar plots for two $p$-type PbTe samples in Figs.( 3(a), 3(e)) for PbTe1 and in Figs. (3(b), 3(f)) for PbTe2. Following procedures enunciated via Fig. 2, we obtain $\Delta_{\rm U} (T)$ and $\Delta_{\rm L}(T)$ for both PbSe and PbS and then plot them in Figs. 4(a) and 4(b). $\Delta_{\rm L}$ depends rather weakly on $T$. In sharp contrast to this, $\Delta_{\rm U}$ grows appreciably with rising $T$, which is consistent with positive temperature coefficients of fundamental band gap found by optical experiments in PbQ \cite{Miller1961, Tauber1966, Tsang1971, Snyder2013}. It is worth mentioning that such a positive rate of change of band gap with $T$ in PbQ is opposite to what occurs in most other semiconductors. In fact, this \emph{anomaly} helps PbQ to achieve high thermoelectric efficiency since it can mitigate the bipolar effects arising from intrinsic carrier activation. The later causes the suppression of thermoelectric figure of merit $zT$ at high $T$'s in a material. We also point out that $\Omega_{\rm C}$ of the $p$-type PbTe sample  can't  be  determined since its conduction band lies in the un-occupied side of its band structure. We can, however, locate $\Omega_{\rm U}$ from Figs. 3(a), 3(e) and $\Omega_{\rm L}$ from Figs. 3(b), 3(f). Therefore, $\Delta (T)$=$\Omega_{\rm U}(T)$-$\Omega_{\rm L}(T)$ of PbTe  like PbSe or PbS can be plotted.

It is evident that $\Delta$ of each PbQ sample decreases monotonically with increasing $T$ (Fig. 4(c)) in the $T$ range of our measurements. $\Delta (T)$ can be well represented by straight lines. From linear extrapolation of $\Delta (T)$ to zero, a characteristic temperature $T^*$  can be defined, at which the BM of the LVB is expected to merge with that of the UVB. Estimated value of  merging temperature $T^*$$\sim$813K (PbTe), $1148$K (PbSe) and $1296$K (PbS). Although such estimation of $T^*$ involves an extrapolation over a large $T$ range, the values of $T^*$ obtained from our ARPES data agree reasonably with those from recent magnetic and optics measurements \cite {Heremans2013, Snyder2013}. We provide further details concerning the connection between various attributes of our $T$ dependent measurements and those from the  literature in the supplementary section. Monotonic $T$ dependence of $\Delta$ suggests that PbQ should become  semiconductors with indirect band gap for $T$$>$$T^*$, where the heavy hole LVB rises in energy above the light hole UVB. In this scenario, the charge transport in PbQ should be dominated by the heavy holes created due to thermal excitations as $T$ approaches $T^*$ and $\Delta$($T$=$T^{*}$)$\sim$$k_{\bf B}T^{*}$. This band convergence increases the density of states of heavier holes, and thus, results in an enhanced Seebeck coefficient and thermoelectric power factor at higher $T$'s. All these are responsible for superior thermoelectric performance of PbQ at elevated temperatures.

\acknowledgments
U.C. acknowledges supports from the National Science Foundation under Grant No. DMR-1454304 and from the Jefferson Trust at the University of Virginia. Work at Argonne National Laboratory (C.D.M., D.Y.C., S.R., M.G.K.) was supported by the U.S. Department of Energy, Office of Basic Energy Sciences, Division of Materials Science and Engineering.

\end{document}